\documentclass[aps,prb,twocolumn,english,showpacs,superscriptaddress,amssymb,amsfonts,nofootinbib]{revtex4-2}
\usepackage{graphicx}
\usepackage[linktocpage]{hyperref}
\hypersetup{colorlinks=true,citecolor=blue,linkcolor=blue, urlcolor=blue, breaklinks=true}
\usepackage{amsmath}
\usepackage{braket}
\usepackage{tikz}
\usetikzlibrary{positioning}
\usepackage{soul,xcolor}
\usepackage[export]{adjustbox}
\usepackage{placeins}
\usepackage[scr=boondox]{mathalpha}
\usepackage{ulem}

\newcommand{\Rom}[1]{\uppercase\expandafter{\romannumeral #1\relax}}
\usepackage{amsthm}

\newtheorem*{assumption*}{\assumptionnumber}
\providecommand{\assumptionnumber}{}


\theoremstyle{definition}

\theoremstyle{theorem}

\theoremstyle{corollary}

\theoremstyle{lemma}

\theoremstyle{Proposition}

\theoremstyle{definition}

\usepackage{amssymb}

\makeatletter

\newcommand{\Rmnum}[1]{\expandafter\@slowromancap\romannumeral #1@}
\makeatother
\newcommand{\imth}{\hspace{1pt}\mathrm{i}\hspace{1pt}}

\newcommand{\bea}{\begin{eqnarray}}
\newcommand{\eea}{\end{eqnarray}}
\newcommand{\bct}{\begin{center}}
\newcommand{\ect}{\end{center}}
\newcommand{\bpm}{\begin{pmatrix}}
\newcommand{\epm}{\end{pmatrix}}
\newcommand{\beq}{\begin{equation}}
\newcommand{\eeq}{\end{equation}}
\newcommand{\bal}{\begin{aligned}}
\newcommand{\eal}{\end{aligned}}
\newcommand{\bfr}{\begin{framed}}
\newcommand{\efr}{\end{framed}}

\newcommand{\expval}[1]{\langle{#1}\rangle}

\begin{document}

\title{Rotational Symmetry Protected Edge and Corner States in Abelian topological phases}
\author{Naren Manjunath}
\affiliation{Department of Physics, Condensed Matter Theory Center, and Joint Quantum Institute, University of Maryland, College Park, Maryland 20742, USA}
\author{Abhinav Prem}
\affiliation{School of Natural Sciences, Institute for Advanced Study, Princeton, NJ 08540, USA}
\author{Yuan-Ming Lu}
\affiliation{Department of Physics, The Ohio State University, Columbus, OH 43210, USA}
	
\date{\today}

\begin{abstract}
Spatial symmetries can enrich the topological classification of interacting quantum matter and endow systems with non-trivial strong topological invariants (protected by internal symmetries) with additional ``weak" topological indices. In this paper, we study the edge physics of systems with a non-trivial shift invariant, which is protected by either a continuous $\text{U}(1)_r$ or discrete $\text{C}_n$ rotation symmetry, along with internal $\text{U}(1)_c$ charge conservation. Specifically, we construct an interface between two systems which have the same Chern number but are distinguished by their Wen-Zee shift and, through analytic arguments supported by numerics, show that the interface hosts counter-propagating gapless edge modes which cannot be gapped by arbitrary local symmetry-preserving perturbations. Using the Chern-Simons field theory description of two-dimensional Abelian topological orders, we then prove sufficient conditions for continuous rotation symmetry protected gapless edge states using two complementary approaches. One relies on the algebraic Lagrangian sub-algebra framework for gapped boundaries while the other uses a more physical flux insertion argument. For the case of discrete rotation symmetries, we extend the field theory approach to show the presence of fractional corner charges for Abelian topological orders with gappable edges, and compute them in the case where the Abelian topological order is placed on the two-dimensional surface of a Platonic solid. Our work paves the way for studying the edge physics associated with spatial symmetries in strongly interacting symmetry enriched topological phases. 
\end{abstract}

	
\maketitle
\tableofcontents

\section{Introduction}
\label{sec:intro}

Symmetry protected topological (SPT) phases offer a rich playground for studying the interplay between symmetry and topology in strongly correlated quantum matter~\cite{chen2012spt,Senthil2015SPT,Chiu2016review}. Theoretically, the classification and characterization of both SPT and symmetry enriched topological (SET) states in two spatial dimensions (2+1D) is well understood in the case when the protecting symmetry $G$ is purely internal, such as $\text{U}(1)_c$ charge conservation, or $\mathbb{Z}_2$ spin-flip symmetries~\cite{chen2012spt,essin2013,mesaros2013,Hung2013,Gu2014Supercoh,kapustin2015fSPT,Lu2016,Tarantino_SET,Barkeshli2019,Wang2020fSPT,barkeshli2021invertible,aasen2021characterization,bulmashSymmFrac,bulmashAnomaly}. However, many physical systems of interest additionally possess spatial symmetries, which play an important role in protecting non-trivial bulk topological properties of 2+1D gapped many-body quantum systems~\cite{cheng2016lsm,Song2017,Thorngren2018,shiozaki2018,you2020hoe,manjunath2021cgt,else2021qc,cheng2022rotation,zhang2022realspace,seiberg2022lsm}. 

Prominent amongst these invariants is the continuum Wen-Zee \textit{shift} $\mathscr{S}$~\cite{Wen1992shift}, which is protected by combined $\text{U}(1)_c$ charge conservation and $\text{SO}(2)\equiv \text{U}(1)_r$ spatial rotation symmetries. A non-trivial Wen-Zee shift manifests as a nonzero Hall viscosity coefficient $\eta_H = \frac{\hbar}{4}\mathscr{S} \bar{n}$, where $\bar{n}$ is the particle number density~\cite{Read2009hvisc}, and as a fractional charge bound to conical defects of the rotational symmetry~\cite{Biswas2016}. Recently, a discrete analog of the Wen-Zee shift was identified in crystalline systems, where the full rotation symmetry is broken to a discrete subgroup~\cite{Han2019,Li2020disc,zhang2022fractional,herzogarbeitman2022rsi}. A non-zero discrete shift manifests, for instance, in the form of fractionally quantized charges at lattice disclinations in the bulk~\cite{zhang2022fractional}; further bulk invariants, including a quantized charge polarization, have also been studied in this context~\cite{zhang2022pol}. Thus, systems with mixed internal and spatial symmetries $G = \text{U}(1)_c \times G_{\text{space}}$ can possess topological invariants in addition to the Chern number, such that even systems with identical Chern numbers can be distinguished through their crystalline topological invariants, which reflect the ``weak" topology of the phase. 

While much progress has been made regarding bulk crystalline indices, the edge physics associated with non-trivial continuum and discrete shift invariants remains less understood, particularly away from the non-interacting limit. This represents a crucial hole in our understanding, since a key experimental signature of topological phases with a bulk gap is the presence of gapless modes localized at edges or corners. For systems with both internal and spatial symmetries, one can also consider interfaces between two phases with identical strong topological invariants but distinct weak topological indices. This suggests the intriguing possibility that an interface between two systems with e.g., identical Chern numbers but distinct crystalline invariants, could host gapless edge states protected by the relevant spatial symmetry, which would provide a crisp, experimentally relevant signature of weak topological indices in quantum many-body systems with non-trivial strong invariants. 

In this paper, we identify the edge manifestations of the continuum and discrete shift invariants in interacting systems, and provide general arguments for their robustness against arbitrary local symmetry preserving perturbations. In the continuum case, we demonstrate the presence of rotation-symmetry protected gapless edge modes at the interface between two quantum Hall systems with identical Chern number but distinct shifts. We provide an analytic argument for this result and supplement it with numerical analysis that supports our conclusions. For systems with intrinsic Abelian topological order, we provide a general understanding of the edge physics using Chern-Simons field theory for SET phases. In the case of discrete rotation symmetries, we use the same field theory approach to show that in systems with a gappable edge, the discrete shift leads to fractional corner charges localized at the vertices of 2d polygons. We further provide a formula to compute the fractional corner charges when any Abelian topological order is placed on the surface of a 3d convex regular polyhedron (a Platonic solid). These results apply to a broad class of gapped quantum many-body phases with charge conservation and rotation (continuous or discrete) symmetries, both with or without intrinsic (Abelian) topological order.

\section{Quantum Hall states in Landau levels}
\label{sec:QHE}

\begin{figure}[b]
    \centering
    \includegraphics[width=0.4\textwidth]{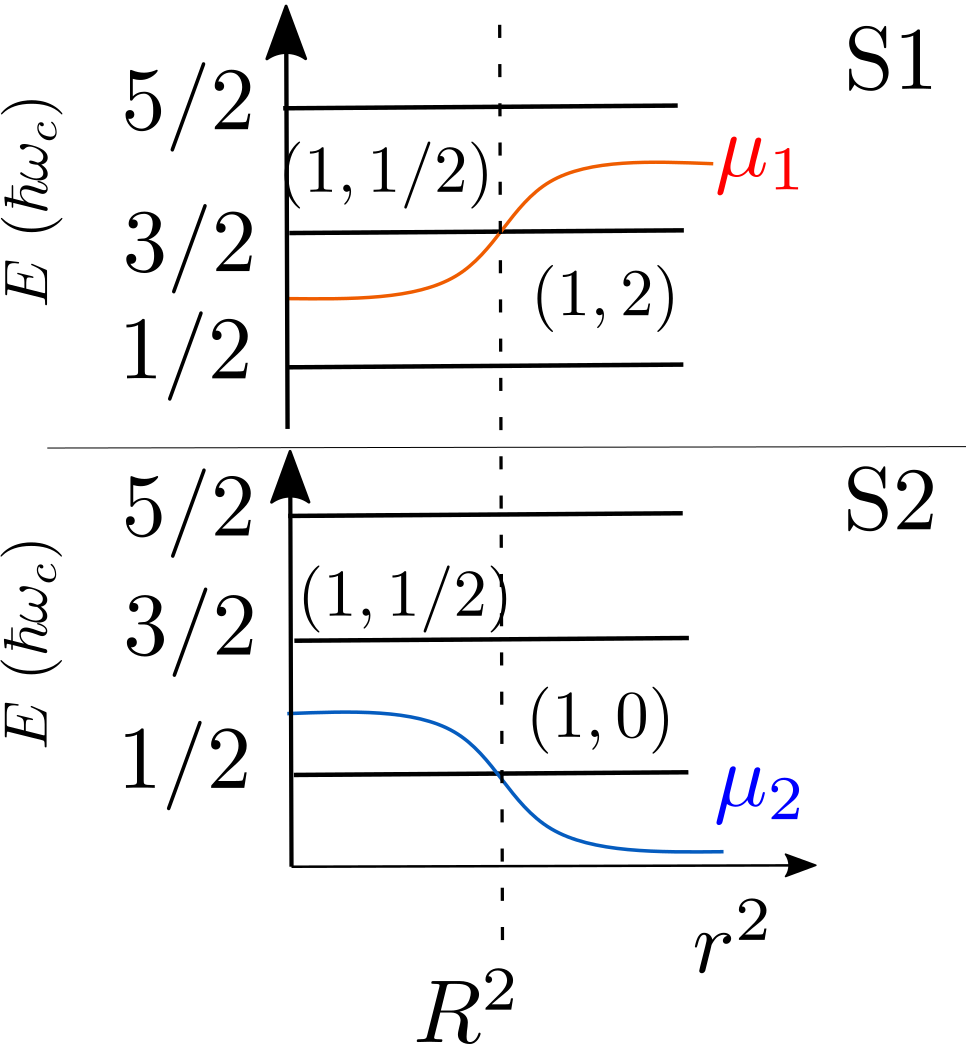}
    \caption{Schematic of a stack of two continuum quantum Hall systems S1, S2 on a disk with chemical potentials $\mu_1,\mu_2$ which are plotted as functions of $r^2$. Black lines correspond to Landau levels without an external potential. The values of $(C,\mathscr{S})$ for each system are written for $r \ll R$ and $r \gg R$.} 
    \label{fig:LLschematic}
\end{figure}

As a concrete example that illustrates our key finding, we consider an interface between two stacks of continuum Landau levels (LLs) such that the Chern number $C$ on either side of the interface is equal while the total Wen-Zee shift $\mathscr{S}$ differs by one. Our construction is described schematically in Fig.~\ref{fig:LLschematic}, where the system has a pair of topological invariants $(C, \mathscr{S}) = (2,1)$ for radius $r \ll R$ (for some fixed $R$), while for $r \gg R$ the invariants are $(C, \mathscr{S}) = (2,2)$. For the first system in the stack, there is a chiral edge mode localized at $r \sim R$ since the Chern number increases by one across the interface; likewise, a \textit{counter-propagating} chiral edge mode results from the second system since its Chern number decreases across the interface. Generically, these edge modes can be gapped by arbitrary local $\text{U}(1)_c$ preserving perturbations. Here, we will show that these counter-propagating zero-energy edge states are in fact protected by rotation symmetry and cannot be gapped by local rotation symmetry preserving perturbations; they are the boundary manifestation of the non-trivial shift invariant. 

Consider a stack of two decoupled quantum Hall systems, each placed on a disc with area $a \gg R^2$ and subject to the same uniform magnetic field $B$. The Hamiltonian for system $i (i = 1,2)$ is
\beq
H_i = \frac{({\bf p}_i + e {\bf A}_i)^2}{2m} -\mu_i({\bf r}_i) \, ,
\eeq
where ${\bf A}_i = \frac{B}{2}(-y_i,x_i,0)$ (symmetric gauge), $\mu_i({\bf r}_i)$ is a slowly varying, radially symmetric chemical potential, and we set $\hbar = c = 1$. The single particle states for each system are given by two harmonic oscillators~\cite{macdonaldreview}:
\begin{equation}
    \ket{n,m}_i := \frac{(a_i^{\dagger})^n (b_i^{\dagger})^m}{\sqrt{n!m!}} \ket{0,0}_i
\end{equation}
where $a_i^{\dagger},b_i^{\dagger}$ are raising operators for the LL index $n$ and another index $m$ respectively, where the angular momentum index $\ell = m -n$. In this representation, the Hamiltonian is
\beq
H = \sum_{i}H_i = \sum_{i} \left[ \left(a_i^\dagger a_i + \frac{1}{2} \right) \omega_c - \mu_i({\bf r}_i) \right] \, ,
\eeq
where $ \omega_c = eB/m$ is the cyclotron frequency. The radial operators $\hat{r}_i^2$ satisfy
\begin{equation}
\label{eq:r2}
    \hat{r}_i^2 = 2 \ell_B^2(1 + b_i^{\dagger} b_i + a_i^{\dagger} a_i - a_ib_i-a_i^{\dagger}b_i^{\dagger})
\end{equation}
where $\ell_B = 1/\sqrt{eB} = 1/\sqrt{\omega_c}$ is the magnetic length; the angular momentum operator is given by
\begin{equation}
\label{eq:Ji}
    J_i = b_i^{\dagger} b_i - a_i^{\dagger} a_i\, .
\end{equation}
These relations are explained further in Appendix~\ref{app:A}. Note that $J = J_1+J_2$ commutes with $H$, since any radial potential conserves angular momentum. 

The chemical potentials
\beq
\mu_i({\bf r}_i) = \frac{2 + (-1)^{i+1}}{2} \omega_c + (-1)^{i+1} K \tanh{\frac{\hat{r}_i^2-R_i^2}{\xi^2}}
\eeq
(with $0<K<\frac{1}{2} \omega_c$ and $R_1,R_2 \approx R \gg \ell_B$) are chosen to ensure that for $r\ll R$ only the lowest LL lies below zero energy in both systems 1 and 2. However, for $r\gg R$, both the $n=0,1$ LLs lie below $E=0$ in system 1, while in system 2 there are no LLs below $E=0$. This implies that upon stacking the two systems, the total Chern number of the filled LLs is $C = 1+1=2$ for $r\ll R$ and $C = 2+0=2$ for $r\gg R$.

We can similarly study the Wen-Zee shift. Using the fact that the shift within the $n$th LL ($n=0,1,2,\dots$) is $n+\frac{1}{2}$~\cite{Wen1992shift}, we see that if $\mathscr{S}^i$ is the shift in system $i$, $\mathscr{S}^1_{r\ll R} = \mathscr{S}^2_{r\ll R} = \frac{1}{2}, \mathscr{S}^1_{r\gg R} = \frac{1}{2}+\frac{3}{2}=2$, and $\mathscr{S}^2_{r\gg R} =0$. Thus, the total shift for the system is $\mathscr{S}_{r\ll R} = 1$ and $\mathscr{S}_{r\gg R} = 2$. Hence, this configuration gives an interface at $r \sim R$ between two systems with identical Chern numbers, but shifts that differ by one.  

\begin{figure}[t]
    \centering
    \includegraphics[width=0.4\textwidth]{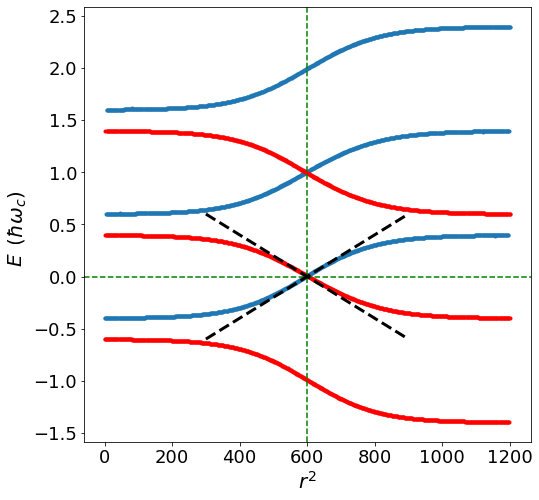}
    \caption{Spectrum of the stacked LL system as a function of $\langle r^2 \rangle$. Red and blue points correspond to states from systems 1 and 2 respectively. Parameters are $m_{\text{max}}=600,R_1^2 = 602, R_2^2 = 598, K = 0.4, \xi^2 = 200$. In this case, $\ell_1 - \ell_2 = -6$. Dashed black lines indicate the approximate linear dispersion for the edge modes (see Eq.~\eqref{eq:approx}), which show excellent agreement with numerics.} 
    \label{fig:LLspectrum}
\end{figure}

Absent any interactions, we expect that the full system must have two zero energy states $\ket{\psi_1},\ket{\psi_2}$ localized at $r \approx R$ with well-defined angular momenta $\ell_1,\ell_2$, where $\ell_i = \bra{\psi_i} J_i \ket{\psi_i}$. This is confirmed in Fig.~\ref{fig:LLspectrum}, which shows the spectrum of the full system (red and blue points correspond to states in systems 1 and 2 respectively). Now, if we found that $\ell_1 = \ell_2$, we would be able to gap out these edge states through rotationally symmetric perturbations. However, we numerically observe that $\ell_1$ cannot be made equal to $\ell_2$ as long as $\ket{\psi_1},\ket{\psi_2}$ are at the same radius i.e., $\bra{\psi_1} \hat{r}_1^2\ket{\psi_1} = \bra{\psi_2} \hat{r}_2^2\ket{\psi_2} = R^2$. This can be achieved by suitably tuning $R_1,R_2$. We note that in the limit where $\xi$ is large, we find that $\ell_1-\ell_2 = -2$; as $\xi$ is decreased, $|\ell_1 - \ell_2|$ increases such that the $\xi \to \infty$ limit provides a lower bound  of 2 on the difference $|\ell_1 - \ell_2|$. As we discuss below, the limiting value has a simple analytic interpretation. Now, if instead $\bra{\psi_1} \hat{r}_1^2\ket{\psi_1} \ne \bra{\psi_2} \hat{r}_2^2\ket{\psi_2}$, the value of $|\ell_1-\ell_2|$ can be changed arbitrarily; however, when we adjust parameters to set $\ell_1 = \ell_2$, we numerically find that the spatial separation between the zero-energy edge states exceeds their localization length ($\sim \ell_B$) so that no \textit{local} rotation symmetric edge perturbation can gap them out (without closing the bulk gap). Moreover, as the potential becomes steeper ($\xi$ decreases), we find that the separation between the zero-energy states increases, such that the matrix element $\bra{\psi_1} V(\hat{r}) \ket{\psi_2}$ for any \textit{local} operator is algebraically suppressed. Thus, our numerical observations support our claim that these zero-energy edge states are robust against arbitrary local rotation symmetric perturbations which do not close the bulk gap.

We can intuitively understand the limiting case as follows. If the chemical potential varies extremely slowly (that is, $\xi$ is very large), there is no mixing between different Landau levels, and so to a good approximation each edge state $\ket{\psi}_i$ has a well-defined value of $n_i^*$ and $m_i^*$, with angular momentum $\ell_i = m_i^* - n_i^*$. 
Now for any state $\ket{n,m}_i$, we have from Eqs~\eqref{eq:r2},\eqref{eq:Ji} that
\begin{align}
\label{eq:r2,Ji}
     _i\bra{n,m} \left( \frac{\hat{r}_i^2}{2\ell_B^2}- J_i -1 \right) \ket{n,m}_i  &= 2n.
\end{align}
Therefore using Eq.~\eqref{eq:r2,Ji}, with $\bra{\psi_i} \hat{r}_i^2 \ket{\psi_i} = R^2$, we obtain
\begin{align}
   \ell_1-\ell_2=\bra{\psi_1} J_1 \ket{\psi_1}-\bra{\psi_2} J_2 \ket{\psi_2} = 2(n_2^*-n_1^*)=-2.  
\end{align}

We can also obtain the same result analytically from the time-independent Schr\"odinger equation. Exploiting the radial symmetry of the problem and writing the wave-function as $\psi(r,\theta) = \frac{u(r) e^{i \ell \theta}}{\sqrt{r}}$, we obtain the radial Schr\"odinger equation in standard WKB form:
\beq
\partial_r^2 u(r) - Q_\ell(r) u(r) = 0 \, ,
\eeq
where 
\beq
Q_\ell(r) = \left(\frac{m^2 \omega_c^2 r^2}{4} + \frac{m \ell \omega_c}{2} - \frac{1/4 - \ell^2}{r^2} -  2m (E + \mu_i(r)) \right) \, .
\eeq
We now consider the limit $R_i \gg \xi \gg \ell_B$, which corresponds to the limit in which the number of flux quanta passing through radius $R_i$ is large: $N_\phi(R_i) = e B \pi R_i^2/2 \pi \gg 1$. Note that this is equivalent to assuming that $\omega_c$ is the largest scale in the system, so we should reproduce the results of the preceding argument. In this limit, a standard WKB analysis~\cite{akkermans1998,cappelli2018} reveals the presence of a linearly dispersing, exponentially localized mode near $r \approx R_i$ for each system:
\begin{align}
\label{eq:approx}
E_{n_0, \ell}^{(i)} &\approx \left(n_0 - \frac{1 + (-1)^{i+1} }{2} \right)\omega_c \nonumber \\
& + (-1)^{i+1} \frac{K R_i^2}{\xi^2}\left(\frac{2n_0 + \ell + 1}{N_\phi(R_i)} - 1 \right) \, .
\end{align}
Since this approximation holds only for $E^{(i)} \ll \omega_c$, we must pick the principal quantum number $n_0 = 1$ ($n_0 = 0$) for the first (second) system, for which this approximate expression shows excellent agreement with numerics (see dashed black lines in Fig.~\ref{fig:LLspectrum}).

Setting $E = 0$, we find the angular momenta $\ell_i$ of the zero-modes
\beq
\label{eq:angmom}
\ell_i = \frac{R_i^2}{2 \ell_B^2} - 2  + (-1)^i \, .
\eeq
Let us first consider the case when $R_1 = R_2$. Then, the difference between the angular momenta of the zero modes is 
\beq
\ell_1 - \ell_2 = -2 \, ,
\eeq
which verifies our claim that these counter-propagating modes cannot be gapped by any rotation-symmetry preserving perturbations since they possess distinct angular momenta, at least when they are localized at the same boundary.

We now consider the case when $R_1 \neq R_2$ and assume that the zero-energy states have identical angular momenta $\ell_1 = \ell_2 = \ell$, for which Eq.~\eqref{eq:angmom} implies $R_1^2 - R_2^2 = 4 \ell_B^2$. Numerically, we observe that this is the minimal amount by which the zero-modes can be separated while equating their angular momenta, which occurs when $\xi$ is made large (this is consistent with the fact that the analytic approximation holds when $\xi \gg \ell_B$). Assuming this minimal separation, if we found that the wave-function overlap between the zero-energy states is non-vanishing, a rotation invariant perturbation would be able to gap them out, contradicting our claim. However, note that the radial component of the zero-mode wavefunctions takes the form~\cite{akkermans1998,cappelli2018}
\beq
\psi^i_{n_0,\ell}(r) \sim \frac{1}{\sqrt{R_i}} H_{n_0}\left(\frac{r - R_i}{\ell_B}\right) \exp\left(-\frac{(r - R_i)^2}{2 \ell_B^2} \right) \, ,
\eeq
(where $H_n(x)$ is the $n$th Hermite polynomial). If the difference in shift were trivial, the two zero-modes would share the same principal quantum number $n^*$; since the overlap between $\psi^1_{n^*,\ell}$ and $\psi^2_{n^*,\ell}$ saturates to an $\mathcal{O}(1)$ constant even in the limit $R_i \gg \ell_B$, these edge states would not be protected. However, in our setup it is crucial that the principal quantum number $n_0$ for the two states differs by one (reflecting the difference in the shift), such that the overlap between $\psi^1_{1,\ell}$ and $\psi^2_{0,\ell}$ (with $R_2^2 - R_1^2 = 4 \ell_B^2$) scales as $\mathcal{O}(\ell_B/R_1)$ in the limit $R_i \gg \ell_B$ (where this analysis holds). Given that a valid edge perturbation is not allowed to close the bulk gap $\sim \omega_c$, this argument supports our claim that the counterpropagating edge states are robust against any local perturbations that respect rotational symmetry.

We have thus provided analytic arguments supported by numerical simulations to demonstrate the existence of rotation-symmetry protected counter-propagating gapless edge modes which are localized at the interface between two quantum Hall systems with identical Chern numbers but distinct shift invariants. We have not yet proven that a nonzero relative shift necessarily implies gapless edge modes; this will be done in the next Section using a field-theory approach that does not rely on the details of any microscopic model. 

The protocol devised here provides a clear, experimentally viable route for probing the (difference in) weak topological indices of systems with identical strong topological invariants. Finally, we note that our results are not restricted to non-relativistic Landau levels -- indeed, we expect that an interface between relativistic Landau levels with identical Chern numbers but distinct shift invariants will also result in rotation-symmetry protected gapless edge states (see e.g. Ref.~\cite{nguyen2021} for a discussion of the Wen-Zee shift in 2+1-D Dirac fermions). 


\section{Gapless edge states in the disc geometry}
\label{sec:gapless}

Motivated by the model study discussed above, we now investigate the general theory of rotation protected edge states in the disc geometry, focusing on the case of Abelian quantum Hall states with both $\text{U}(1)_c$ charge conservation and a continuous spatial rotational symmetry $\text{U}(1)_r$. In the framework of Abelian Chern-Simons theory and edge chiral boson edge states~\cite{Wen1995}, we shall derive and prove sufficient conditions for gapless edge states protected by rotation symmetry $\text{U}(1)_r$. We work on a disc geometry to ensure compatibility with $\text{U}(1)_r$. Here, we will not consider discrete translation symmetry, although we expect our discussion can be straightforwardly generalized to include the additional weak invariants that are protected by $\mathbb{Z}^2$ translations~\cite{manjunath2021cgt,zhang2022pol}.

\subsection{Field theory of edge states in the disc geometry}

We consider a generic two-dimensional (2+1D) Abelian topological order enriched with both the charge conservation symmetry $\text{U}(1)_c$ and a continuous spatial rotational symmetry $\text{U}(1)_r \simeq \text{SO}(2)$. This theory is described by the following multi-component Abelian Chern-Simons theory~\cite{wen1992kMatrix}:
\bea\notag
&\mathcal{L}_\text{bulk}=-\frac{\epsilon^{\mu\nu\lambda}}{4\pi}K_{I,J}a_\mu^I\partial_\nu a^J_\lambda\\
&+ \frac{\epsilon^{\mu\nu\lambda}}{2\pi}t_IA_\mu\partial_\nu a^I_\lambda+\frac{\epsilon^{i\mu\nu}}{2\pi}s_I\omega_i\partial_\mu a_\nu^I\label{eq:bulk action}
\eea
where we follow the Einstein convention to always sum over repeated indices. $K_{I,J}$ is an $N_K\times N_K$ integer-valued symmetric matrix, which is invertible for a gapped topological order. $A$ and $\omega$ are gauge fields for the $\text{U}(1)_c$ and $\text{U}(1)_r$ symmetries respectively.\footnote{The field theory treats the spatial rotation symmetry $\text{U}(1)_r$ effectively as an internal symmetry. This assumption is discussed further in e.g. Refs.~\cite{Thorngren2018,manjunath2022constraint}.} $\vec t$ and $\vec s$ are known as the charge and spin vectors of the Abelian topological order, characterizing the charge and angular momentum carried by quasiparticles in the topological order~\cite{Wen1995}. $t_I$ are all integers, while $s_I$ can be either integers or half-integers in bosonic or fermionic systems respectively~\cite{Wen1992shift}. 

Upon integrating out the gauge fields $a^I$, we obtain the following effective response theory~\cite{Gromov2015,manjunath2021cgt}:
\begin{align}
    \mathcal{L}_{\text{eff}} &= \frac{\sigma_{xy}}{4\pi}\epsilon^{\mu\nu\lambda} A_{\mu}\partial_{\nu} A_{\lambda} + \frac{\mathscr{S}}{2\pi}\epsilon^{\mu\nu\lambda} A_{\mu}\partial_{\nu} \omega_{\lambda} \nonumber \\&+ \frac{\ell_s}{4\pi}\epsilon^{\mu\nu\lambda} \omega_{\mu}\partial_{\nu} \omega_{\lambda},\\
    \sigma_{xy} &= t^T K^{-1} t; \quad \mathscr{S} = s^T K^{-1} t; \quad \ell_s = s^T K^{-1} s.
\end{align}
Here, we have ignored the contribution from the `framing anomaly' discussed in Ref.~\cite{Gromov2015}. 

We now place the system on a disc of radius $R$, preserving the $\text{U}(1)_r$ rotational symmetry. Quantizing the above Chern-Simons theory on an open disc $\mathcal{D}$ of radius $R$ leads to a constraint $\epsilon^{ij}\partial_ia_j^I=0$ in the bulk, and hence $a_j^I=\partial_j\phi^I$. This leads to the following effective Lagrangian density on the circular edge parametrized by the coordinate $x=R\theta$, with $\theta\in[0,2\pi]\simeq S^1$:
\bea\notag
\mathcal{L}_\text{edge}=-\frac{K_{I,J}}{4\pi}\partial_x\phi^I\partial_t\phi^J+\frac{t_I\epsilon^{\mu\nu}}{2\pi}A_\mu\partial_\nu\phi^I+\frac{s_I}{2\pi}\omega_x\partial_t\phi^I+\cdots\\
\label{eq:edge action}
\eea
where $\cdots$ represent the non-universal energetic terms. Hereafter, we shall use the polar angle $\theta$ and the edge coordinate $x=\theta R$ interchangeably to parametrize the edge states on the disc of radius $R$. The chiral bosons satisfy the following commutation relation:
\bea\label{eq:kac moody alg}
[\phi^I(x),\partial_y\phi^J(y)]=-2\pi\imth\delta(x-y)K^{-1}_{I,J} \, .
\eea
The charge density on the edge is given by
\bea\label{charge density}
\rho_c=\frac{t_I}{2\pi}\partial_x\phi^I \, ,
\eea
and similarly the spin angular momentum density on the edge is written as
\bea
\rho_{s}=\frac{s_I}{2\pi}\partial_x\phi^I \, .
\eea
Under a $\text{U}(1)$ charge rotation by phase $\alpha$, a generic edge excitation $\hat V_{\vec l}\sim e^{\imth l_I\phi^I(x,t)}$ transforms as
\bea\label{sym:U(1)}
e^{\imth\alpha\int\rho(x)\text{d}x}V_{\vec l}e^{-\imth\alpha\int\rho(x)\text{d}x}=e^{-\imth\alpha l_IK^{-1}_{I,J}t_J}V_{\vec l} \, .
\eea
In other words, the above edge excitation $V_{\vec l}$ carries a $\text{U}(1)_c$ charge of
\bea
Q_{\vec l}=-l_IK^{-1}_{I,J}t_J \, .
\eea
Similarly, the $\text{U}(1)_r$ spatial rotation is generated by the total angular momentum
\bea\label{angular momentum}
L_z=-\imth\partial_\theta+\int\rho_s(x)\text{d}x \, .
\eea
Under a spatial rotation by angle $\alpha$, the edge excitation $\hat V_{\vec l}$ transforms as 
\bea\label{sym:rot}
e^{\imth\alpha L_z}V_{\vec l}(x)e^{-\imth\alpha L_z}=e^{-\imth\alpha l_IK^{-1}_{I,J}s_J}V_{\vec l}(x+R\alpha) \, .
\eea
In other words, a rotational-invariant edge excitation
\bea
V_{\vec l,n}\equiv\int_0^{2\pi}\text{d}\theta e^{-\imth n\theta}V_{\vec l}(x=\theta R)
\eea
carries angular momentum
\bea
L_z(\vec l,n)=n-l_IK^{-1}_{I,J}s_J \, .
\eea

Below, we present and derive three sufficient conditions for gapless edge states in the disc geometry, protected by $\text{U}(1)_c\times \text{U}(1)_r$ symmetries. We derive each condition using two different approaches: the first one is based on the sufficient and necessary conditions for a gapped open boundary introduced in Ref.~\cite{levin2013}, while the second derivation is based on the Lieb-Schultz-Mattis-Oshikawa type flux insertion argument~\cite{OshikawaLSM}. 

\subsection{Sufficient conditions for gapless edge states}

In the absence of any symmetry, in Ref.~\cite{levin2013} Levin established the following theorem regarding the robustness of edge excitations of 2+1D Abelian topological orders, as described by Eq.~\eqref{eq:edge action} (see also Ref.~\cite{haldane1995}). The edge states of a 2+1D Abelian topological order described by Chern-Simons theory (Eq.~\eqref{eq:bulk action}) with matrix $K$ can be gapped if and only if there exists a \emph{Lagrangian subgroup}~\cite{Kapustin2011,Davydov2013} $\mathcal{M}=\{m_i\}$ of integer vectors $m_i$, defined by the following two conditions:

(i) $m_i^TK^{-1}m_j=0\mod1,~~~\forall~m_i,m_j\in\mathcal{M}$;

(ii) For any quasiparticle labeled by an integer vector $\vec l$, it either satisfies $\vec l\in\mathcal{M}$ (i.e. $\vec l=\sum_ic_i m_i$), or has nontrivial braiding statistics with at least one element in $\mathcal{M}$ (i.e. $\exists~m_i\in\mathcal{M}$ such that $m_i^TK^{-1}l\neq0\mod1$). 

Ref.~\cite{levin2013} also provided the explicit form of the backscattering terms (also called Higgs terms~\cite{moroz2017}) that gap out the edge states if both conditions are satisfied\footnote{Here, $f(x)$ is any periodic function which is smooth along any smooth part of the boundary but can have discontinuities at corners when only $\text{C}_n$ rotation symmetry is present.}:
\bea\label{gap term}
\mathcal{L}_\text{b.s.}=\sum_{\{\Lambda_i\}}U_{\Lambda_i}(x)\cos(\Lambda_i^IK_{I,J}\phi^J-f_i(x))
\eea
where the \emph{null vectors} $\{\Lambda_i\}$ satisfy $\Lambda_i^TK\Lambda_j=0,~\forall~i,j$~\cite{levin2013}. The physical meaning of an element $\vec m_i\in\mathcal{M}$ of the Larangian subgroup is a bosonic quasiparticle $V_{\vec m}\sim e^{\imth m^I\phi^I}$ which condenses on the gapped edge, in the sense that the edge operator $V_{\vec m}$ has a long-range ordered correlation function on the gapped edge~\cite{levin2013}:
\bea\label{edge correlator}
\lim_{|x-y|\rightarrow\infty}\expval{e^{\imth m_I\phi^I(x)}e^{-\imth m_I\phi^I(y)}}\neq0,~~~\forall~\vec m\in\mathcal{M.}
\eea
While the above theorem applies to edge states without any symmetry, below we consider how the presence of continuous $\text{U}(1)_c\times \text{U}(1)_r$ symmetries adds new constraints for obtaining a symmetrically gapped edge. 

First of all, we consider the $\text{U}(1)_c$ symmetry associated with charge conservation. In this case, in order for the gapped edge to preserve $\text{U}(1)_c$ symmetry, any charged operator must have a short-ranged correlation function, since otherwise the $\text{U}(1)_c$ symmetry is spontaneously broken. Assuming a gapped symmetric edge, due to the property Eq.~\eqref{edge correlator}, we must have
\bea\label{eq:charge U(1):constraint}
m^TK^{-1}t=0,~~~\forall~m\in\mathcal{M} 
\eea
otherwise there will be long-range ordered correlations for the $\text{U}(1)_c$-charged operator $e^{\imth m_i^I\phi^I}$, a signature for the spontaneous breaking of the charge $\text{U}(1)_c$ symmetry. We now show that a fractional Hall conductance
\bea\label{cond:charge}
\sigma_{xy}=t^TK^{-1}t\neq0\mod1
\eea
is a sufficient condition for gapless edge states. We consider an edge excitation $V_{\vec t}$ associated with the charge vector $\vec t$. By definition (i) of the Lagrangian subgroup $\mathcal{M}$, the fractional Hall conductance Eq.~\eqref{cond:charge} dictates that $\vec t\notin\mathcal{M}$. On the other hand, the relation Eq.~\eqref{eq:charge U(1):constraint} states that $\vec t$ has trivial braiding with all elements of $\mathcal{M}$. This contradiction to condition (ii) of a Lagrangian subgroup suggests that a gapped edge preserving $\text{U}(1)$ symmetry is impossible. Meanwhile, a continuous $\text{U}(1)$ symmetry cannot be spontaneously broken in one spatial dimension. Thus, Eq.~\eqref{cond:charge} is a sufficient condition for gapless edge modes in 2+1D Abelian topological orders (Eq.~\eqref{eq:bulk action}). 

It is now straightforward to generalize the above arguments to $\text{U}(1)_r$ spatial rotational symmetry. In order for the long-range order of $V_{\vec m_i}$ operators (or more precisely, $\expval{V^\dagger_{\vec m_i,n_i=0}V_{\vec m_i,n_i=0}}\neq0$) to not spontaneously break $\text{U}(1)_r$ rotational symmetry (Eq.~\eqref{sym:rot}), we must require that
\bea
m^TK^{-1}s=0,~~~\forall~m\in\mathcal{M}
\eea
Now, following the same argument as in the case of $\text{U}(1)_c$ symmetry, we find
\bea\label{cond:rot}
s^TK^{-1}s\neq0\mod1
\eea
as another sufficient condition for gapless edge states in the disc geometry. 

Finally, in the presence of the full $\text{U}(1)_c\times \text{U}(1)_r$ symmetry, assuming a gapped symmetric edge, in order for both $\vec t$ and $\vec s$ to satisfy condition (ii) of a Lagrangian subgroup, it is straightforward to show that
\bea\label{cond:charge+rot}
\mathscr{S}=t^TK^{-1}s\neq0\mod 1.
\eea
Thus, a fractional Wen-Zee shift $\mathscr{S}$ is also a sufficient condition for gapless edge states in the disc geometry. 

\subsection{Flux insertion arguments}

Above, using the necessary and sufficient conditions for a gapped edge without any symmetry~\cite{levin2013}, we showed that the presence of $\text{U}(1)_c\times \text{U}(1)_r$ symmetry gives rise to extra necessary conditions for obtaining a gapped symmetric edge. In particular, we derived the sufficient conditions Eqs.~\eqref{cond:charge},\eqref{cond:rot}, and~\eqref{cond:charge+rot} for gapless edge states in the disc geometry, protected by the $\text{U}(1)_c\times \text{U}(1)_r$ symmetry. In this derivation, we required (\ref{cond:charge}), (\ref{cond:rot}), and (\ref{cond:charge+rot}) to have a nonzero fractional part, relying on the braiding statistics argument of Ref.~\cite{levin2013}, which only detects fractional statistics. Below, we provide an alternative proof based on the flux insertion argument~\cite{OshikawaLSM}, which allows us to expand the sufficient conditions and simply requires (\ref{cond:charge}), (\ref{cond:rot}), or (\ref{cond:charge+rot}) to be nonzero.

As a warm up exercise, we first derive a well known result: the edge states of a quantum Hall state must be gapless if it has a nonzero Hall conductance~\cite{laughlin1981,Halperin1982,LevinStern2012} i.e.
\bea\label{cond:charge:gauging}
\sigma_{xy}=t_IK^{-1}_{I,J}t_J\neq0 \, .
\eea
To prove this conclusion, we first assume a gapped symmetric edge, and then use the flux insertion argument to derive a contradiction. If the edge states can be gapped out without symmetry breaking, there will be a unique many-body ground state in the disc geometry, separated from the rest of the spectrum by a finite energy gap. The finite gap allows us to adiabatically thread a total $\text{U}(1)$ flux of $\Phi$ through the disc, e.g. uniformly over the bulk of the disc, without closing the gap. Now that the low energy subspace of the whole system is effectively spanned by the edge excitations described by Eq.~\eqref{eq:edge action}, we focus on how the adiabatic flux insertion process influences the edge states. To be specific, we assume that the edge states are symmetrically gapped out by adding the following generic local terms to the Lagrangian density:
\bea\notag
&\delta\mathcal{L}_\text{edge}(x=\theta R)=\sum_{\{\vec l_i\}}V_{K\vec l_0}(\theta_0=\theta)\cdot\\
&\int\prod_{i=1}^{N_v}\Big[\frac{\text{d}\theta_i}{2\pi}V_{K\vec l_i}(\theta+\theta_i)\Big]T_{\{\vec l_i\}}(\{\theta_i\})\label{edge ham}
\eea
While $\text{U}(1)_c$ symmetry Eq.~\eqref{sym:U(1)} imposes the constraint
\bea
\sum_il_i^Tt=0 \, ,
\eea
the $\text{U}(1)_r$ rotational symmetry Eq.\eqref{sym:rot} requires
\bea
\sum_il_i^Ts=0 \, .
\eea
Adiabatic insertion of uniform flux $\Phi$ (in units of $\frac{\Phi_0}{2\pi}=\hbar/e$) in the bulk leads to a vector potential of $A_\theta=\Phi/2\pi$ on the circular edge of the disc, which modifies the edge Hamiltonian Eq.~\eqref{edge ham} as
\bea
\tilde T_{\{\vec l_i\}}(\{\theta_i\})=e^{\imth A_\theta\sum_{i=1}^{N_v}\theta_i l_{i}^Tt} T_{\{\vec l_i\}}(\{\theta_i\}) \, .
\eea
After the $\Phi=2\pi$ flux insertion, the above change to the edge Hamiltonian can be absorbed by the following large gauge transformation~\cite{Lieb1961,OshikawaLSM}:
\bea
U_0=e^{\imth\int\text{d}xt_I\phi_I(x)} \, .
\eea
In other words, the $2\pi$-flux-inserted Hamiltonian $H(\Phi=2\pi)$ is related to the original zero-flux Hamiltonian through
\bea
U_0H(\Phi=2\pi)U_0^{-1}=H(\Phi=0) \, .
\eea
In the presence of a finite energy gap, the adiabatic flux insertion process relates the unique ground states of $H(\Phi = 0)$ and $H(\Phi = 2\pi)$ by the large gauge transformation, up to a phase $e^{\imth\alpha}\in \text{U}(1)$:
\bea
\ket{\Phi=0}=e^{\imth\alpha}U_0\ket{\Phi=2\pi} \, .
\eea
However, note that the total $\text{U}(1)_c$ charge Eq.~\eqref{charge density} on the edge does not commute with $U_0$:
\bea
U_0\big(\int\rho_c(x)\text{d}x\big)U_0^{-1}=\int\rho_c(x)\text{d}x-t^TK^{-1}t \, .
\eea
In other words, the total $\text{U}(1)_c$ charge of the edge ground state changes under the flux insertion process, contradicting our assumption of a unique gapped ground state on the edge. Therefore, it is impossible to have a gapped symmetric edge with a nonzero Hall conductance (\ref{cond:charge:gauging}).

We can now straightforwardly generalize this argument and similarly relax the constraint in Eq.~\eqref{cond:charge+rot}. Note that the total angular momentum Eq.~\eqref{angular momentum} does not commute with $U_0$ either:
\bea
U_0L_zU_0^{-1}=L_z-s^TK^{-1}t \, .
\eea
Therefore, if
\bea\label{cond:charge+rot:gauging}
\mathscr{S} =s^TK^{-1}t\neq0
\eea
it is impossible to have a gapped edge that preserves both $\text{U}(1)_c$ and $\text{U}(1)_r$ symmetry. This mixed anomaly of $\text{U}(1)_c\times \text{U}(1)_r$ symmetry provides another sufficient condition for gapless edge states on a disc geometry. 

Unlike the adiabatic insertion of $\text{U}(1)_c$ flux which can be implemented both in microscopic lattice models and in the continuum field theory as shown above, to insert the $\text{U}(1)_r$ flux of the spatial rotational symmetry, one needs to create conical defects~\cite{Biswas2016}; in practice, this is subtle to carry out in a microscopic lattice model. However, the continuum field theory Eq.~\eqref{eq:edge action} of the edge states permits us to conveniently insert a $\text{U}(1)_r$ flux $\Phi_r$\footnote{Although the $\text{U}(1)_r$ and $\text{U}(1)_c$ symmetry are treated on equal footing as internal symmetries in the bulk effective field theory Eq.~\eqref{eq:bulk action}, their fluxes have different manifestations on the edge states. In particular, the spatial $\text{U}(1)_r$ flux includes a rescaling of the angle variable $\theta$ in contrast to the global $\text{U}(1)_c$ flux.}, after which the modified edge Hamiltonian Eq.~\eqref{edge ham} becomes 
\bea
\tilde T^\prime_{\{\vec l_i\}}(\{\theta_i\})=e^{\frac{\imth\Phi_r}{2\pi}\sum_{i=1}^{N_v}\theta_i l_{i}^Ts} T_{\{\vec l_i\}}\big(\{(1+\frac{\Phi_r}{2\pi})\theta_i\}\big)
\eea
We assume a gapped symmetric edge, when both the chiral central charge $c_-$ and the charge Hall conductance vanish, and adiabatically insert a $2\pi$ flux of $U(1)_r$ symmetry. As before, the adiabatic insertion of $2\pi$ flux can be absorbed by the following large gauge transformation:
\bea
U_0^\prime=e^{\theta\partial_\theta+\imth\int\text{d}xs_I\phi_I(x)}
\eea
Since the total angular momentum Eq.~\eqref{angular momentum} of the edge states is not preserved during the flux insertion process:
\bea
U_0^\prime L_z(U_0^\prime)^{-1}=L_z-s^TK^{-1}s \, ,
\eea
it is impossible to have a gapped edge preserving $\text{U}(1)_r$ symmetry. We have hence proved the third sufficient condition 
\bea\label{cond:rot:gauging}
s^TK^{-1}s\neq0
\eea
for gapless edge states in the disc geometry, protected by the spatial rotation symmetry $\text{U}(1)_r$.

\section{Discrete rotational symmetry and corner charges}

In this section, we investigate geometries beyond the disc geometry discussed above. In particular, we compute the corner charges of Abelian topological orders on a 2d regular polygon, and on the 2d surface of a three-dimensional (3d)  regular polyhedron (i.e. a Platonic solid). Note that while corner charges have mostly been discussed in the context of free-fermion higher-order topological insulators in 2d and 3d~\cite{Benalcazar2017,Benalcazar2019HOTI,Liu2019,Schindler2020,Watanabe2021corner,takahashi2021corner,naito2022corner,MayMann2022HOTI,you2020hoe,Jiang2022}, our results apply more generally in the presence of strong interactions.


First, we discuss an Abelian topological order (given by Eq.~\eqref{eq:bulk action}) on a regular polygon of $n$ sides (i.e. an $n$-gon), where the continuous spatial rotational symmetry $\text{U}(1)_r$ is broken down to a discrete $n$-fold rotation $\text{C}_n$. If the topological order has a gappable edge~\cite{levin2013} with vanishing charge and thermal Hall conductance ($c_-=\sigma_{xy}=0$), the gapless edge states protected by conditions (\ref{cond:charge+rot:gauging}) or (\ref{cond:rot:gauging}) can be gapped out on the $n$ open edges of an $n$-gon. Nevertheless, such a $\text{C}_n$-symmetric $n$-gon will exhibit corner charges. To be specific, the following back-scattering term (see Eq.~\eqref{gap term})
\bea\notag
&H_{b.s.}=\sum_{\{\Lambda_i\}}U_{\Lambda_i}\int_0^{2\pi}\text{d}\theta\cos\big[\Lambda_i^TK\phi(\theta)-f_i(\theta)\big],\\
&f_i(\theta+2\pi/n)=f_i(\theta)+\frac{2\pi}{n}\Lambda_i^Ts
\eea
can gap out the edge modes while preserving the $\text{C}_n$ symmetry, where we have chosen $f_i(\theta)$ to be constants along each edge which jumps across each corner. The $\text{C}_n$ symmetry can be verified using the transformation rule Eq.~\eqref{sym:rot}. As usual, we use the polar angle $\theta\in[0,2\pi]\simeq S^1$ to label the coordinate of the chiral boson fields $\{\phi^I(\theta)\}$ on the open edges of the $n$-gon. Using the commutation relations Eq.~\eqref{eq:kac moody alg} and transformation under rotational symmetry Eq.~\eqref{sym:rot}, it is straightforward to show that each corner (at $\theta=\theta_0$) of the $n$-gon is associated with the following vertex operator:
\bea
\mathcal{D}_{\text{C}_n}(\theta_0)\sim e^{\imth s_I\phi^I(\theta_0)/n} \, .
\eea
Using the $\text{U}(1)_c$ symmetry transformation Eq.~\eqref{sym:U(1)}, we then find that the corner charge is given by\footnote{Here, we do not consider the effect of discrete translation symmetries, which can contribute additional terms to the effective response theory; see e.g.~\cite{manjunath2021cgt,zhang2022pol}.}
\bea\label{corner charge}
Q_{\text{C}_n}=-\frac1{n}t^TK^{-1}s = -\frac{\mathscr{S}}{n}\mod1 \, .
\eea

\begin{table}[t]
\begin{tabular}{ |c|c|c|c|c|} 
 \hline
Platonic solids &$n$&$m$&$V$&$Q_{(n,m)}$\\
 \hline
Tetrahedron&3&3&4&$\mathscr{S}/2$\\
\hline
Cube&4&3&8&$\mathscr{S}/4$\\
\hline
Octahedron&3&4&6&$\mathscr{S}/3$\\
\hline
Dodecahedron&5&3&20&$\mathscr{S}/10$\\
\hline
Icosahedron&3&5&12&$\mathscr{S}/6$\\
\hline
\end{tabular}
\caption{The 5 distinct Platonic solids in three dimensions, and the fractional charge Eq.~\eqref{vertex charge} localized on each vertex when a 2+1D topological order (given by Eq.~\eqref{eq:bulk action}) is placed on their 2d surfaces.}\label{table:platonic solids}
\end{table}

Next, we consider the case where the 2+1D Abelian topological order is placed on the 2d surface of a 3d convex regular polyhedron (also known as a Platonic solid). The faces of each Platonic solid are congruent regular $n$-gons, and each vertex is shared by $m$ faces. The five distinct Platonic solids are summarized in Table~\ref{table:platonic solids}, where the number $F$ of faces, $V=nF/m$ of vertices, and $E=nF/2$ of edges satisfy the Euler characteristic of $V-E+F=2$ for 3d convex polyhedrons. Therefore the vertex number is given by
\bea
V=\frac{4n}{2(m+n)-mn}
\eea
In a Platonic solid, each vertex is joined by $m$ corners of $n$-gons. As detailed in Appendix \ref{app:corner charge}, a direct calculation based on concrete edge gapping terms leads to the following fractional charge accumulated on each vertex
\bea\label{vertex charge}
Q_{(n,m)}=\frac{2(m+n)-mn}{2n}\mathscr{S}\mod1
\eea
This is consistent with the prediction of the Wen-Zee term\cite{Wen1992shift}, since each corner of the Platonic solid has a Gaussian curvature of $4\pi/V$, the charge accumulated at the corner should be given by
\bea
Q_{(n,m)}=\frac{4\pi}{V}\frac{\mathscr{S}}{2\pi}=\frac{2(m+n)-mn}{2n}\mathscr{S}\mod1
\eea
the same as produced by the direct calculation. Note that Eq.~\eqref{corner charge} only applies for 2+1D topological orders with a gappable edge~\cite{levin2013}, whose charge and thermal Hall conductance must vanish, so that each side of the 2d $n$-gon can be gapped out. In contrast, Eq.~\eqref{vertex charge} applies to a \textit{generic} 2+1D topological order irrespective of whether it has a gapped edge or not. 


\section{Concluding Remarks}
\label{sec:cncls}

In summary, we have studied the boundary excitations of interacting quantum phases with arbitrary 2+1D Abelian topological orders that preserve both charge $\text{U}(1)_c$ and spatial rotation $\text{U}(1)_r$ symmetries, for different geometries. In the disc geometry, which preserves the continuous $\text{U}(1)_r$ symmetry, we derived three sufficient conditions (\ref{cond:charge:gauging}), (\ref{cond:charge+rot:gauging}), and (\ref{cond:rot:gauging}) for gapless edge states using Chern-Simons field theory. We also demonstrated these results explicitly through a microscopic calculation for quantum Hall states in Landau levels, where we showed the presence of rotation symmetry protected gapless edge modes on the interface between two systems with identical Chern numbers but distinct Wen-Zee shifts. In the 2d regular polygon geometry, where the $\text{U}(1)_r$ symmetry is broken down to a discrete $\text{C}_n$ rotational symmetry, we further derived the general formula Eq.~\eqref{corner charge}) for the fractional charge bound to each corner, under the condition of a gappable edge with vanishing Hall condutance. Using this result, we computed the fractional charge Eq.~\eqref{vertex charge} bound to each vertex of a Platonic solid, when an arbitrary Abelian topological order is placed on the 2d surface of the Platonic solid. 

While we have focused on Abelian quantum Hall states in this work, one natural future direction is to generalize it to non-Abelian topological orders. Moreover, in the case of discrete $\text{C}_n$ rotational symmetry, the Wen-Zee term~\cite{Wen1992shift,Han2019} considered in this work is only applicable to the cases where $\text{C}_n$ symmetry does not permute anyons in the topological order. When distinct anyons are transformed into each other by the $\text{C}_n$ symmetry, more exotic non-Abelian corner states can emerge~\cite{Clarke2013,Barkeshli2013genon,Lu2014,Teo2015}. More broadly, our work opens the door for studying interfaces between strongly interacting systems (either with or without intrinsic topological order) which possess the same strong SPT invariants (such as the Chern number) but distinct weak invariants (such as the shift). We expect that similar results can be obtained for the remaining weak invariants (such as the charge polarization) that appear in the formal classification of 2+1D topologically ordered systems with $\text{U}(1)_c$ charge conservation, $\mathbb{Z}^2$ translation, and $\text{C}_n$ discrete rotation symmetries~\cite{manjunath2021cgt}. We leave a thorough investigation of the edge manifestations of the remaining weak topological indices to future work. 


\section*{Acknowledgments}

We are particularly grateful to Barry Bradlyn for comments and feedback on the draft. N.M. thanks Yuxuan Zhang, Gautam Nambiar, and Maissam Barkeshli for discussions and collaboration on related projects. A.P. thanks Jonah Herzog-Arbeitman for stimulating discussions on related subjects. We thank the Simons Center for Geometry and Physics for its hospitality, where this work was initiated and partially carried out during the program ``Geometrical Aspects of Topological Phases of Matter: Spatial Symmetries, Fractons and Beyond.'' N.M. and A.P. gratefully acknowledge hospitality and support from the Institute for Advanced Study, where part of this work was carried out. YML thanks KITP for hospitality, where this research was supported in part by NSF Grant No. PHY-1748958 and the Gordon and Betty Moore Foundation Grant No. 2919.02. This work is supported by the Laboratory for Physical Sciences through the Condensed Matter Theory Center (N.M.), by the U.S. Department of Energy, Office of Science, Office of High Energy Physics under Award Number DE-SC0009988 (A.P.), by NSF under Award No. DMR 2011876 (Y.-M.L.) and DMR- 1753240 (N.M.). 
\newline
\paragraph*{Note added:}  During the completion of this work, we became aware of a related work Ref.~\cite{Rao2023} which considers the boundary response of higher-order topological insulators with $\text{C}_4$ rotation symmetry. Our work agrees where it overlaps.



\bibliographystyle{apsrev}
\bibliography{refs}


\appendix


\section{Background on the numerical calculation}
\label{app:A}

In the main text, we studied a stack of two LL systems with radially varying potentials. Here, we explain how to model one such system, following the procedure delineated in Ref.~\cite{macdonaldreview}. Note that the Hamiltonian for noninteracting fermions in a background magnetic field described by the vector potential ${\bf A}$ and a radial potential $V(r)$ is
\begin{equation}
    H = \frac{({\bf p}-e {\bf A})^2}{2m} + V(r) \, ,
\end{equation}
where ${\bf A}$ is chosen to be in the symmetric gauge. We first express this Hamiltonian in the Landau quantized picture for a general choice of $V$.

This is done by rewriting $H$ in terms of two independent harmonic oscillator degrees of freedom. Let $\hat{a},\hat{a}^{\dagger}$ be the operators that lower/raise the LL index $n$. Let $\hat{b},\hat{b}^{\dagger}$ be the operators that lower/raise a second quantum number $m$. The angular momentum in this representation is given by $\ell=m-n$. In the symmetric gauge, each eigenstate (when $V=0$) can be written in the form $\ket{n,m} := \frac{(a^{\dagger})^n (b^{\dagger)^{m}}}{\sqrt{m! n!}}\ket{0,0}$. Then we have
\begin{align}
    a \ket{n,m} &= \sqrt{n}\ket{n-1,m} \\
    a^{\dagger} \ket{n,m} &= \sqrt{n+1}\ket{n+1,m} \\
    b \ket{n,m} &= \sqrt{m}\ket{n,m-1} \\
    b^{\dagger} \ket{n,m} &= \sqrt{m+1}\ket{n,m+1}.
\end{align}

In this basis, the kinetic term of $H$ is simply
\begin{equation}
    \hbar \omega_c \left( a^{\dagger} a + \frac{1}{2} \right).
\end{equation}
Now, we want to express the radial potential $V$ in this basis as well. Note that $\hat{x},\hat{y}$ can be written in terms of these operators as
\begin{align}
    \frac{\hat{x} + i \hat{y}}{\sqrt{2}\ell_B} &= i (a - b^{\dagger}) \\
   \frac{\hat{x} - i \hat{y}}{\sqrt{2}\ell_B} &= -i (a^{\dagger} - b).
\end{align}
This implies that
\begin{equation}
    \hat{r}^2 = 2 \ell_B^2(1+ a^{\dagger} a + b^{\dagger} b - a b - a^{\dagger} b^{\dagger} ).
\end{equation}
In particular, 
\begin{align}
    \frac{\hat{r}^2}{\ell_B^2} \ket{n,m} &= 2(n+m) \ket{n,m} \nonumber \\ &- 2 \sqrt{mn} \ket{n-1,m-1} \nonumber \\ &- 2 \sqrt{(n+1)(m+1)} \ket{n+1,m+1}.
\end{align}
We see that $\hat{r}^2$ conserves the angular momentum $\ell = m-n$. We can now express $\hat{r}^2$ in the basis of $\ket{n,m}$, diagonalize it, and obtain a set of eigenstates $\ket{\phi_j}, j = 1,2, \dots, $ with eigenvalues $r^2_j$ (which we assume increase with the index $j$). Let $\Phi$ be the eigenvector matrix associated to $\hat{r}^2$, and $D$ be the diagonal matrix with $D_{ii} = r_i^2$. Then, for a general radial potential, we can write $V(\hat{r}^2) = \Phi V(D) \Phi^{-1}$. This gives us the matrix for the desired $V$ in the $\ket{n,m}$ basis.

\section{Computing the corner charge on a Platonic solid}\label{app:corner charge}

\begin{figure}[t]
    \centering
    \includegraphics[width=0.4\textwidth]{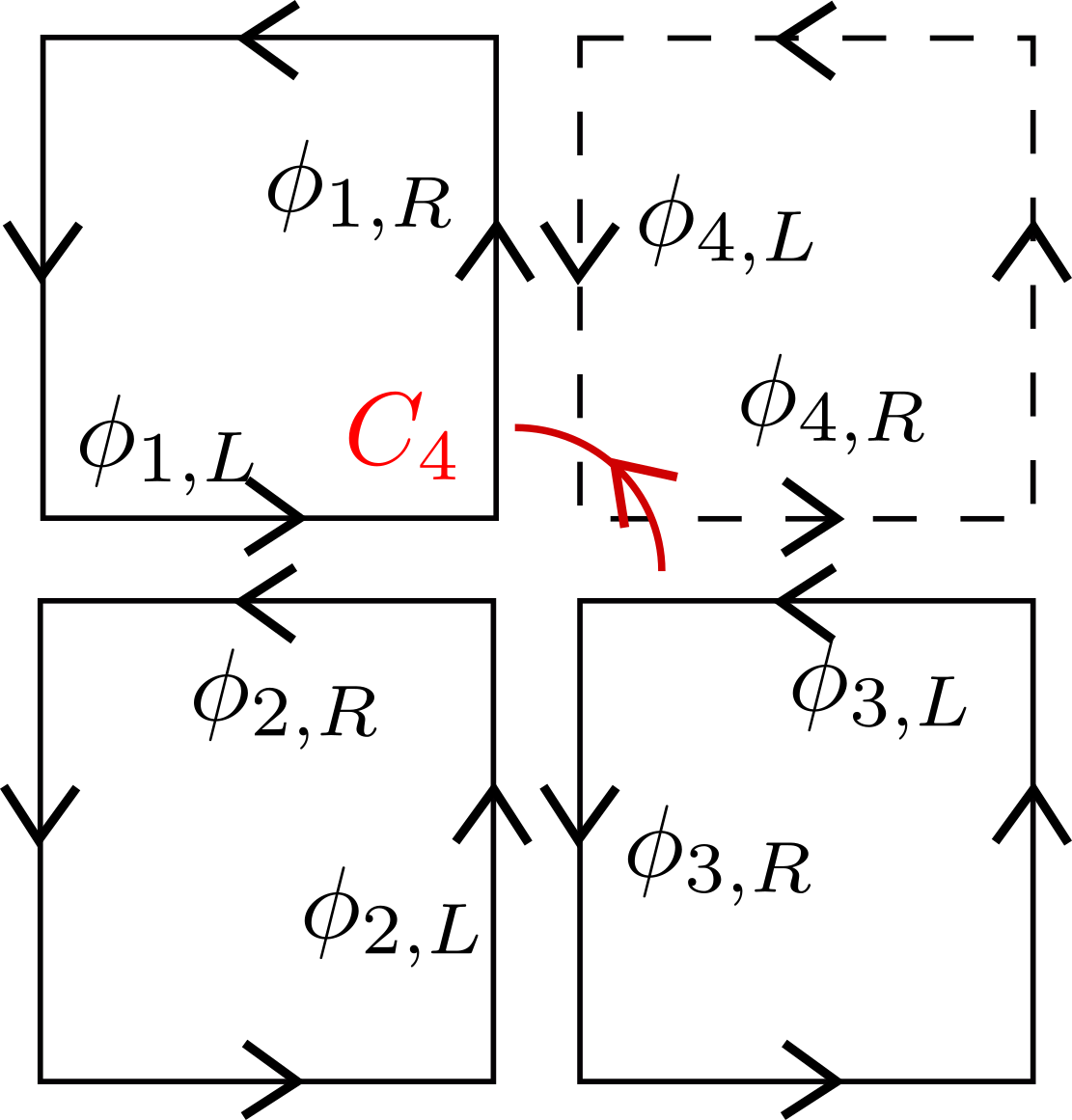}
    \caption{An illustration of how a 2d square-shaped plane can be by sewing 4 squares together around one vertex. If only 3 squares are sewed together around the vertex (neglecting the dashed square), one corner of a 3d cube is formed.}
    \label{fig:edge_gap}
\end{figure}

Before putting the 2+1-D topological order on the surface of a 3d regular polyhedron (i.e. a Platonic solid), we start by considering a flat 2d geometry, where 
\bea\label{def:N}
N=\frac{2\pi}{(n-2)\pi/n}=\frac{2n}{n-2}
\eea
$n$-gons join at a vertex to form a flat 2d system, since every interior angle of the $n$-gon is $\theta_n=\frac{(n-2)\pi}{n}$. The $n=N=4$ case is illustrated in Fig. \ref{fig:edge_gap}. We use the vector $\vec\phi\equiv(\phi_1,\cdots,\phi_{N_K})^T$ to label the gapless chiral bosons on the edge of the Abelian topological order. By labeling the chiral bosons $\{\vec \phi_{j,L/R}|1\leq j\leq N\}$ and on the pair of edge in each $n$-gon, the following gapping term describe the gapped bulk of the flat 2d system:
\bea\label{gap term:2d plane}
\mathcal{H}_\text{gap}^0=-\sum_{\{e_i\}}\sum_{j=1}^NU_{e_i}\cos\big[e^T_iK(\phi_{j+1,R}-\phi_{j,L})\big]
\eea
where we have chosen the vectors $\{(e_i)_I=\delta_{i,I}|1\leq i\leq N_K\}$ to be a complete basis of the $N_K$-dimensional integer lattice. Physically, it pins the chiral bosons to the following minimum of energy:
\bea
\expval{\phi_{j+1,R}}=\expval{\phi_{j,L}}
\eea
on each edge, so that all anyons (represented by vertex operator $\hat V_{\vec l}\sim e^{\imth l_I\phi^I(x,t)}$ in terms of chiral bosons on the edge) can freely tunnel through the edge from one $n$-gon to another. Note that under the $N$-fold rotational symmetry $C_N$ around the vertex where $N$ $n$-gons join, the edge chiral bosons transform as 
\bea
C_N \phi_{j,L/R}C_N^{-1}= \phi_{j+1,L/R}-\frac{(n-2)\pi}{n}K^{-1}s
\eea
due to (\ref{sym:rot}). As a result, with the gapping Hamiltonian (\ref{gap term:2d plane}), the bosons are pinned to the following $C_N$-symmetric minimum:
\bea
\expval{\phi_{j+1,L/R}}=\expval{\phi_{j,L/R}}+\frac{(n-2)\pi}{n}K^{-1}s
\eea
Meanwhile, the integral of the local charge density (\ref{charge density}) across the corner of each $n$-gon leads to the total charge localized at the vertex:
\bea
&Q=\sum_{j=1}^N\frac{t^T}{2\pi}\expval{\phi_{j,R}-\phi_{j,L}}\notag\\
&\equiv\sum_{j=1}^N\frac{t^T}{2\pi}\expval{\phi_{j+1,R}-\phi_{j,L}}=0\mod1
\eea
consistent with a smooth bulk in a flat 2d system.

Now we are ready to discuss the case of a Platonic solid, where $m$ (instead of $N$ as discussed previously) $n$-gons join at each corner to form a corner. The $n=4,~m=3$ case is again illustrated in Fig. \ref{fig:edge_gap}. The first $m-1$ terms of the gapping Hamiltonian (\ref{gap term:2d plane}) remains the same, while an extra term sews the two edges described by $\phi_{m,L}$ and $\phi_{1,R}$. Note that in the previous case of a flat 2d system, we have 
\bea
\expval{\phi_{1,R}}=\expval{\phi_{N,L}}=\expval{\phi_{m,L}}+(N-m)\frac{(n-2)\pi}{n}
\eea
due to the rotational symmetry $C_N$ around the vertex (i.e. the corner here). In order to properly sew the edge states without breaking the rotational symmetry, the new gapping term for each corner of a Platonic solid is
\bea
&\notag\mathcal{H}_\text{gap}=-\sum_{\{e_i\}}U_{e_i}\Big\{\sum_{j=1}^{m-1}\cos\big[e^T_iK(\phi_{j+1,R}-\phi_{j,L})\big]\\
&+\cos\big[e^T_iK(\phi_{1,R}-\phi_{m,L}-(N-m)\frac{(n-2)\pi}{n})\big]\Big\}\label{gap term:corner}
\eea
As a result, the corner charge of a 2d Abelian topological order on a Platonic solid is given by
\bea
&Q_{n,m}=\sum_{j=1}^m\frac{t^T}{2\pi}\expval{\phi_{j,R}-\phi_{j,L}}\notag\\
&\equiv\sum_{j=1}^m\frac{t^T}{2\pi}\expval{\phi_{j+1,R}-\phi_{j+1,L}}\notag\\
&=\frac{(N-m)(n-2)}{2n}t^TK^{-1}s\mod1\notag\\
&=\frac{2(m+n)-mn}{2n}t^TK^{-1}s\mod1
\eea
where we have used relation (\ref{def:N}).


\end{document}